\documentclass[a4paper, superscriptaddress, amsfonts, amssymb, amsmath, reprint, showkeys, nofootinbib]{revtex4-2}
\usepackage[english]{babel}
\usepackage[utf8]{inputenc}
\usepackage[colorinlistoftodos, color=green!40, prependcaption]{todonotes}
\usepackage{amsthm}
\usepackage{mathtools}
\usepackage{physics}
\usepackage{xcolor}
\usepackage{graphicx}
\usepackage[left=23mm,right=13mm,top=35mm,columnsep=15pt]{geometry} 
\usepackage{adjustbox}
\usepackage{placeins}
\usepackage[T1]{fontenc}
\usepackage{lipsum}
\usepackage{csquotes}
\usepackage{floatrow}
\usepackage{float}

\usepackage{nicefrac}
\usepackage{amsfonts}

\usepackage{amssymb}
\usepackage{amsmath} 
\usepackage[caption=false]{subfig}
\usepackage{multirow} 
\usepackage{tabularx} 
\usepackage{array}

\usepackage{units}

\usepackage{tensor} 
\usepackage{braket}

\usepackage{bm}
\usepackage{hyperref}

\newcommand{\RomanNumeralCaps}[1]
    {\MakeUppercase{\romannumeral #1}}
\newcommand{\angstrom}{\mbox{\normalfont\AA}}

\graphicspath{{}}
\begin{document}

\title{Quantum Hall effect in InAsSb quantum wells at elevated temperatures}

\author{M.E. Bal}
\email[Correspondence email address: ]{maurice.bal@ru.nl}
\affiliation{High Field Magnet Laboratory (HFML-EMFL), Radboud University, Toernooiveld 7, 6525 ED Nijmegen, The Netherlands} 
\affiliation{Institute for Molecules and Materials, Radboud University,Heyendaalseweg 135, 6525 AJ Nijmegen, The Netherlands}
\author{E. Cheah}
\affiliation{Solid State Physics Laboratory, ETH Zurich, CH-8093 Zurich, Switzerland}
\affiliation{Quantum Center, ETH Zürich, CH-8093, Switzerland}
\author{Z. Lei}
\affiliation{Solid State Physics Laboratory, ETH Zurich, CH-8093 Zurich, Switzerland}
\affiliation{Quantum Center, ETH Zürich, CH-8093, Switzerland}
\author{R. Schott}
\affiliation{Solid State Physics Laboratory, ETH Zurich, CH-8093 Zurich, Switzerland}
\affiliation{Quantum Center, ETH Zürich, CH-8093, Switzerland}
\author{C. A. Lehner}
\affiliation{Solid State Physics Laboratory, ETH Zurich, CH-8093 Zurich, Switzerland}
\author{H. Engelkamp}
\affiliation{High Field Magnet Laboratory (HFML-EMFL), Radboud University, Toernooiveld 7, 6525 ED Nijmegen, The Netherlands} 
\affiliation{Institute for Molecules and Materials, Radboud University,Heyendaalseweg 135, 6525 AJ Nijmegen, The Netherlands}
\author{W. Wegscheider}
\affiliation{Solid State Physics Laboratory, ETH Zurich, CH-8093 Zurich, Switzerland}
\affiliation{Quantum Center, ETH Zürich, CH-8093, Switzerland}
\author{U. Zeitler}
\email[Correspondence email address: ]{uli.zeitler@ru.nl}
\affiliation{High Field Magnet Laboratory (HFML-EMFL), Radboud University, Toernooiveld 7, 6525 ED Nijmegen, The Netherlands} 
\affiliation{Institute for Molecules and Materials, Radboud University,Heyendaalseweg 135, 6525 AJ Nijmegen, The Netherlands}

\date{\today} 

\begin{abstract}
We have characterized the electronic properties of a high-mobility two-dimensional electron system in modulation doped InAsSb quantum wells and compare them to InSb quantum wells grown in a similar fashion. Using temperature-dependent Shubnikov-de Haas experiments as well as FIR transmission we find an effective mass of $m^{*} \approx$ 0.022$m_{e}$, which is lower than in the investigated InSb quantum well, but due to a rather strong confinement still higher than in the corresponding bulk compound. The effective $g$-factor was determined to be $g^{*} \approx$ 21.9. These results are also corroborated by \textit{\textbf{k $\cdot$ p}} band structure calculations. When spin polarizing the electrons in a tilted magnetic field, the $g$-factor is significantly enhanced by electron-electron interactions, reaching a value as large as $g^{*}$ = 60 at a spin polarization $P$ = 0.75. Finally, we show that due to the low effective mass the quantum Hall effect in our particular sample can be observed up to a temperature of 60 K and we propose scenarios how to increase this temperature even further.
\end{abstract}

\maketitle

\section{Introduction}
Since its discovery in 1980 by von Klitzing \textit{et al.}, the quantum Hall effect (QHE) has been observed in a variety of high-mobility 2D systems, ranging from MOSFETs \cite{klitzing1980new}, quantum wells (QWs) \cite{stormer1979electronic,PhysRevB.86.045404} as well as intrinsic 2D materials \cite{novoselov2005two,bandurin2017high,li2016quantum,tang2021two}. This resulted in the definition of the resistance quantum, which enabled an incredibly precise determination of the fine-structure constant. Another breakthrough was achieved with the discovery of room temperature (RT) QHE in graphene, making it a convenient platform for performing metrological experiments \cite{novoselov2007room,Small_1997,Hartland1987TheRB,jeffery1997nist,jeckelmann2001quantum,janssen2012precision}. Since the robustness of the QHE is, amongst others, governed by the Landau level (LL) spacing ($\Delta_{LL}$ $\sim$ 2000 K at 30 T for graphene), InSb with its small effective mass ($m_{bulk}^*$ = 0.014$m_{e}$), would conceptually also be a good material for the observation of a RT QHE. Optimally alloying InSb with InAs can reduce the effective mass even further, by utilizing the band bending commonly observed in ternary \RomanNumeralCaps{3}-\RomanNumeralCaps{5} compound semiconductors \cite{stradling1997magnetotransport,vurgaftman2001band}. The QHE in InSb QWs has been investigated from several perspectives, e.g. FQHE \cite{PhysRevResearch.4.013039,PhysRevResearch.2.033213}, spin effects \cite{PhysRevResearch.4.013039,PhysRevResearch.2.033213}, high-current breakdown \cite{PhysRevB.86.045404} and Ising quantum Hall ferromagnetism \cite{PhysRevB.69.235315}, whereas the literature on InAsSb QWs is less abundant. However, this material has become the subject of renewed interest because of its strong spin-orbit interaction \cite{PhysRevB.106.165304,PhysRevB.108.L121201} and small direct band gap \cite{svensson2019temperature}. \par 
In this paper, we present magneto-transport and far infrared (FIR) transmission data of InSb and InAsSb modulation doped QWs, enabling a direct comparison between the parent compound and the alloy. Firstly, the temperature dependence of Shubnikov-de Haas (SdH) oscillations enables the determination of the effective mass. Simultaneously, we acquire the temperature where the QHE disappears. The results for the effective mass are consistent with those obtained from cyclotron resonance (CR) measurements. Coincidence experiments in tilted magnetic fields are used to extract the \textit{g}-factors, which are consistent with band-structure calculations based on \textit{\textbf{k $\cdot$ p}} theory. Additionally, an interaction-induced enhancement of the \textit{g}-factor for low filling factors, i.e. high spin polarization, is observed. 

\section{Samples and Methods}
Both the InSb as well as the InAsSb QW discussed in this manuscript, were grown by molecular beam epitaxy (MBE). The sample structure is based on the work of Lehner \textit{et al.} \cite{20.500.11850/336239}, who systematically investigated the effect of employing buffer layers to overcome the lattice mismatch between (100) GaAs substrates and the InSb QWs \cite{PhysRevMaterials.2.054601}. On the other hand, the InAsSb QW was grown on a (001) GaSb substrate, thereby removing the AlSb/GaSb intermediate transition. The underlying In$_{1-y}$Al$_{y}$Sb/In$_{1-x}$Al$_{x}$Sb (x=0.42, y=0.54) buffer was adjusted to the InAsSb QW lattice constant. The InSb (InAs$_{0.38}$Sb$_{0.62}$) 21 nm thick QW was surrounded by an Al$_{0.1}$In$_{0.9}$Sb (Al$_{0.42}$In$_{0.58}$Sb) confinement barrier, which contained an asymmetric double-sided Si $\delta$-doping layer, located 30 nm (44 nm) above and below the QW. The layer stack of both QW structures, as well as an investigation of the InAsSb QW and its As/Sb ratio via XRD, TEM, and EDX, are shown in Fig. 1 and 2 of the supplemental information \cite{supplemental}.\par
For the magneto-transport measurements both heterostructures were processed into a 880 $\times$ 25 $\mu m^{2}$ Hall bar using wet chemical etching. Ar-milling the top surface removes any contaminants and ensures that the successive layers of Ge/Ni/Au evaporated onto it, provide high quality Ohmic contacts. More information about the microfabrication can be found in Refs.~\onlinecite{lei2019quantum,PhysRevResearch.2.033213}. The FIR transmission measurements were performed with an unprocessed piece of the same wafers, warranting a better signal-to-noise compared to the Hall bar. In order to suppress Fabry-P\'erot interference effects in the transmitted light, the back of the sample was wedged under an angle of $\sim$4$^{\circ}$.\par
The transport measurements were performed in a $^4$He bath cryostat, containing a variable temperature insert (VTI) with T = 1.4 - 300 K, which was placed inside a 38 T Florida-Bitter magnet. By mounting the sample on a rotating platform, it was possible to change the angle of the magnetic field \textit{in-situ}. Standard low-frequency lock-in techniques were used to acquire $\rho_{xx}$ and $\rho_{xy }$ of a 310 $\times$ 25  (405 $\times$ 25) $\mu m^{2}$ section of the InAsSb (InSb) Hall bar. \par 

\begin{figure}[t]
\centering
\includegraphics[width=8.4cm,height=7.6cm]{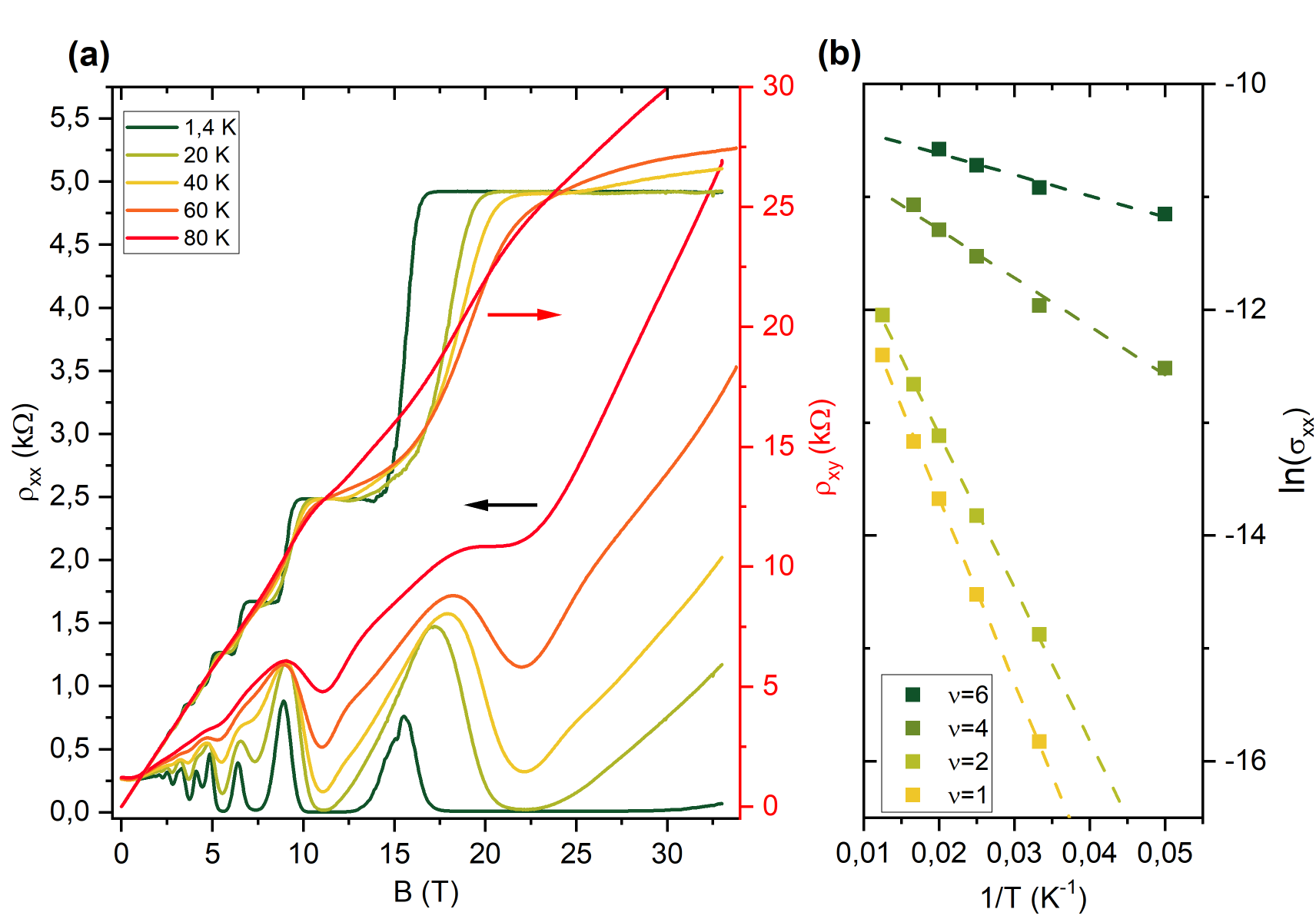}
\caption{(a) Magnetic field dependence of $\rho_{xx}$ and $\rho_{xy}$ in an InAsSb QW at different temperatures. (b) Temperature dependence of $\sigma_{xx}$ at multiple filling factors. Using the relation $\sigma_{xx} \propto exp(-E_{act}/2k_{B}T)$, we extracted the activation energies from the linear fits.}
\label{fig:1}
\end{figure}

A Fourier transform infrared (FT-IR) spectrometer (Bruker VERTEX 80v) with a Globar broadband source, was used for the FIR transmission experiment. The IR radiation was coupled into a quasi-optical beamline, which guided the radiation towards a 33 T Florida-Bitter magnet. At the end of the beamline the radiation was focused into an oversized brass waveguide and passed through a polyethylene window, into a new-silver waveguide that was inside a $^4$He bath cryostat. Before reaching the sample, the radiation was focused using a brass cone. The transmitted IR radiation was collected by a second brass cone, passed through another polyethylene window onto a Si bolometer that was kept at superfluid helium temperatures. The FT-IR spectra were measured at different fixed magnetic fields and normalized with a background spectrum, which was constructed from all spectra by taking the 90$^{th}$ percentile of each frequency, to obtain the transmittance $T(B) = T_{m}(B)/T_{ref}$.

\section{Results}
In the following section only the InAsSb QW will be discussed, though comparisons will be made between these results and those obtained with the InSb QW. The results from the InSb can be found in the supplemental information (SI).

\begin{figure}[t]
\centering
\includegraphics[width=8.4cm,height=7.6cm]{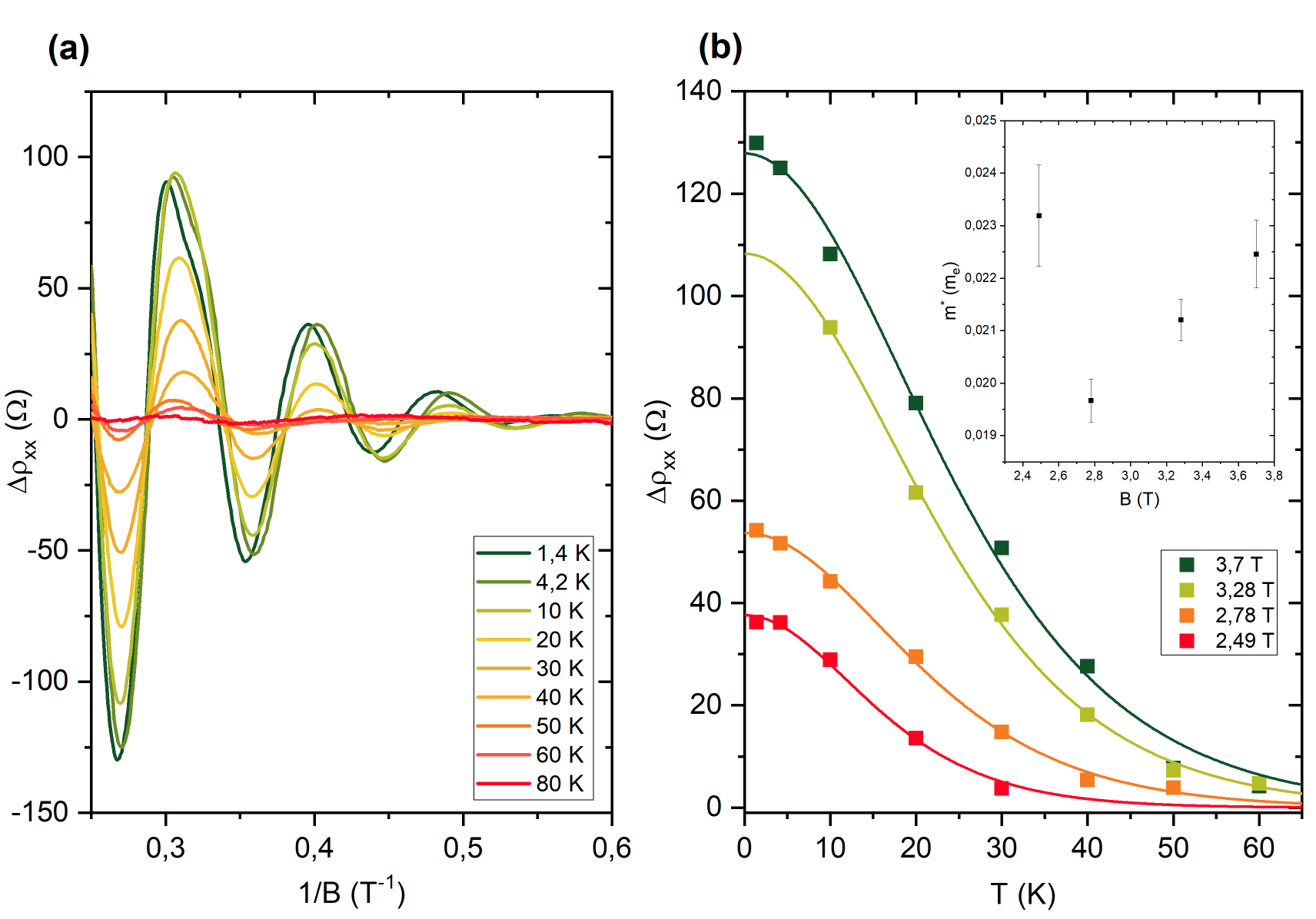}
\caption{(a) Temperature dependence of the SdH oscillations in an InAsSb QW after subtraction of a magnetoresistance background. Shown is only the field range were no spin splitting is observed. (b) Temperature dependence of quantum oscillation amplitudes at different magnetic fields. The solid lines represent fits to Eq. 1 and the effective mass deduced from them is depicted in the inset.}
\label{fig:2}
\end{figure}
 
\subsection{SdH and quantum Hall measurements}

As illustrated in Fig. \ref{fig:1}a, the magneto-transport measurements performed on a InAsSb QW Hall bar show clear SdH oscillations in $\rho_{xx}$ and quantized plateaus in $\rho_{xy}$ at $h/\nu e^{2}$, where $\nu$ is the filling factor. A carrier concentration of $n$ = 5.39 $\times$ 10$^{11}$ cm$^{-2}$ and $n$ = 4.67 $\times$ 10$^{11}$ cm$^{-2}$ was extracted from the classical Hall effect in small magnetic fields for the InAsSb and InSb QW, respectively. The corresponding mobility $\mu$ for both the InAsSb and InSb QW were 4.24 $\times$ 10$^{4}$ cm$^{2}$V$^{-1}$s$^{-1}$ and 3.37 $\times$ 10$^{5}$ cm$^{2}$V$^{-1}$s$^{-1}$. In an InAsSb QW at the lowest temperatures we can resolve $\nu$ = 12 and $\nu$ = 8 in $\rho_{xx}$ and $\rho_{xy}$, respectively. This is considerably lower than for our high-mobility InSb QW where $\nu$ = 42 was observed. As the condition $\mu B \gg 1$ must be met in order to observe quantum oscillations, one would expect the onset of SdH oscillations to be at $B \sim$ 2.4 T. This prediction is in good agreement with the experimental data presented in Fig. \ref{fig:1}a. On the other hand, spin-splitting occurs at a magnetic field of 3.2 T corresponding to $\nu$ = 7. This difference may be attributed to alloy disorder scattering, which is intrinsic for ternary compounds, or a possible alloy inhomogeneity in the QW \cite{PhysRevB.106.165304}. In Fig. \ref{fig:1}b, we extract the quantum Hall activation energies of different filling factors, which will be discussed in greater detail in the upcoming section. \par

\begin{figure}[b]
\centering
\includegraphics[width=8.4cm,height=7cm]{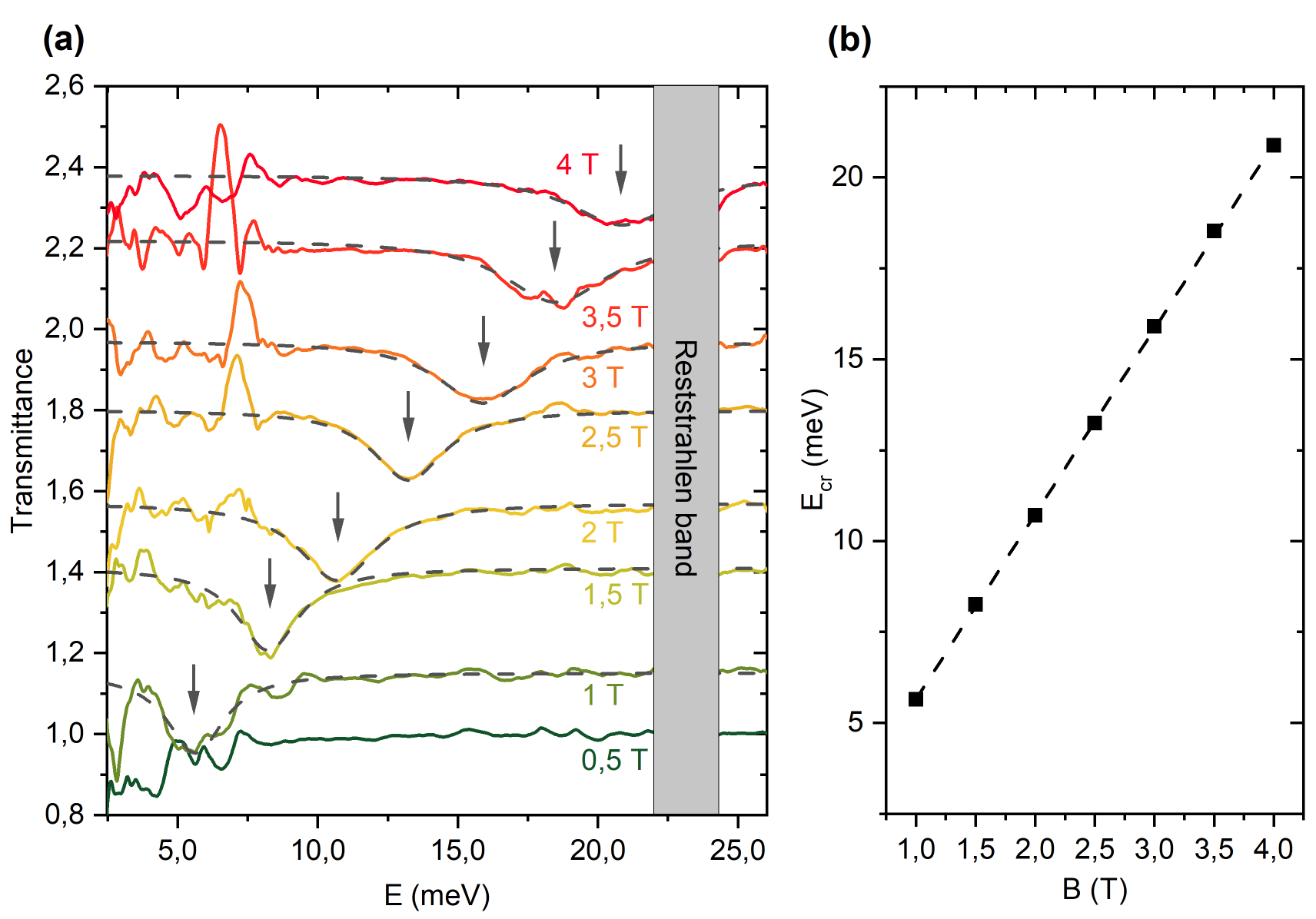}
\caption{(a) Transmittance as a function of energy for different magnetic fields in an InAsSb QW. The traces have a constant offset of 0.2. In the region between 3-6 meV the signal is fairly noisy due to lower signal strength and the pick up of electronic/mechanical noise. Due to the Reststrahlen band, the sample becomes completely opaque between 22-24 meV \cite{Li_1992}. The arrows indicate the position of the CR determined from the Lorentzian fits (gray dashed line). (b) Cyclotron resonance energy as function of magnetic field. }
\label{fig:7}
\end{figure}

\begin{figure}[t]
\centering
\includegraphics[width=8.4cm,height=6cm]{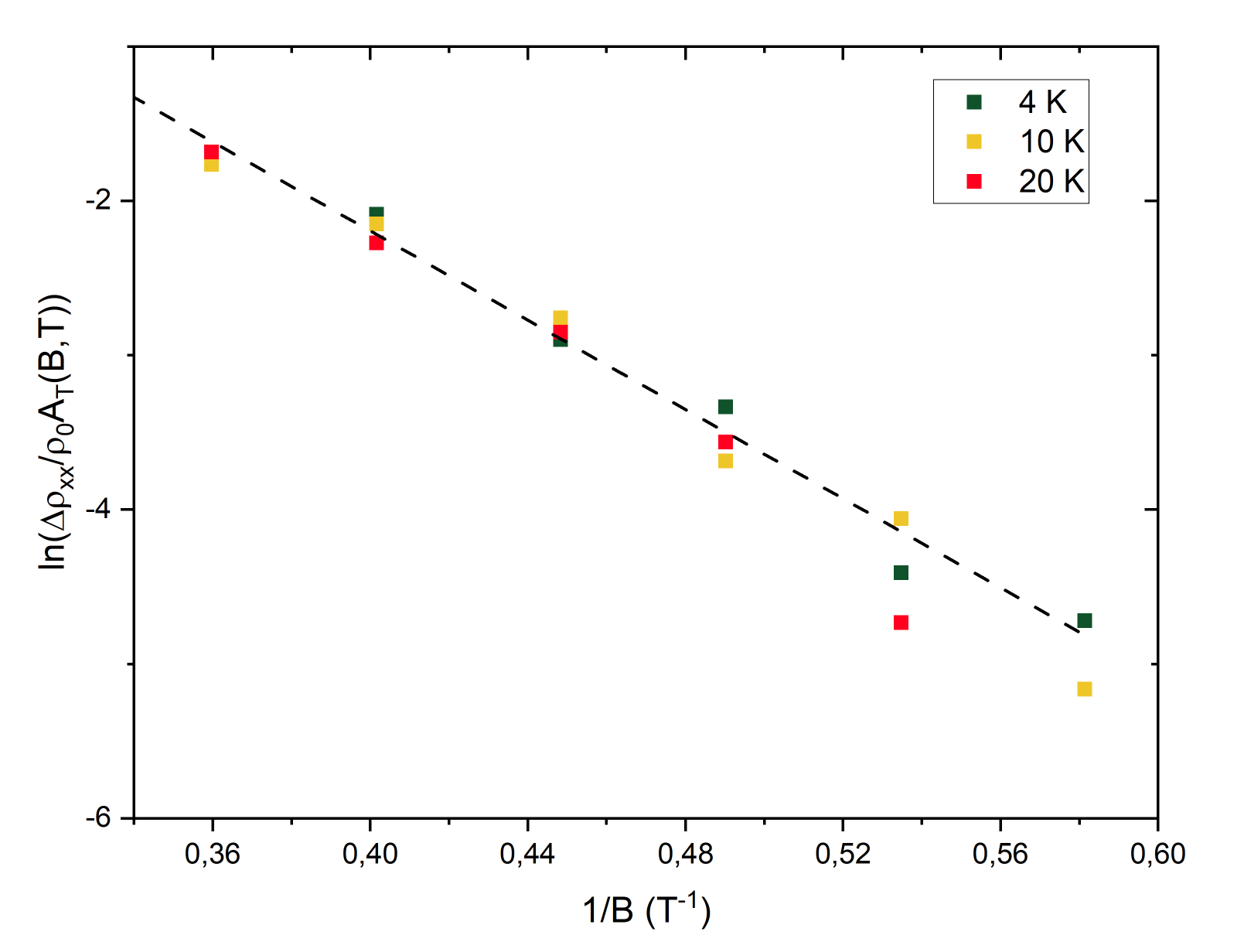}
\caption{Quantum lifetime determined from the linear fit of the exponential Dingle term in the Ando formula for the Shubnikov-de Haas effect in an InAsSb QW.}
\label{fig:9}
\end{figure}

\subsection{Effective mass, g-factor and lifetime analysis}

We start our investigation of the effective mass by introducing the Ando's expression for the quantum oscillation amplitude in 2D systems \cite{miura2007physics}
\begin{equation}
    \frac{\Delta\rho_{xx}(B,T)}{\rho_{0}(B)}\propto A_{T}(B,T)A_{D}(B)A_{s},
\end{equation}
where 
\begin{equation}
    A_{T}(B,T) = \frac{\frac{2C\pi^{2}k_{B}T}{\hbar\omega_{c}}}{\sinh(\frac{2\pi^{2}k_{B}T}{\hbar\omega_{c}})},
\end{equation} 
\begin{equation}
    A_{D}(B)=\exp(\frac{-\pi}{\omega_{c}\tau_{q}}),
\end{equation} 
\begin{equation}
    A_{s}=\cos{\bigg[2\pi\bigg(\frac{g^{*}m^{*}}{2m_{e}}-\frac{1}{2}\bigg)\bigg]},
\end{equation}

$\omega_{c} = eB/m^{*}$ is the cyclotron frequency and $C$ is a constant related to the oscillation amplitude at 0 K. In order to extract the effective mass from the temperature dependent suppression of the SdH oscillations, we first subtract a smooth magneto-resistance background $\rho_{0}$ from the $\rho_{xx}$ data shown in Fig. \ref{fig:1}a, giving us the actual quantum oscillation amplitude $\Delta\rho_{xx}$ (see Fig. \ref{fig:2}a). Figure \ref{fig:2}b shows the suppression of some local maxima and minima of $\Delta\rho_{xx}$ with increasing temperature, which was fitted with $A_{T}(B,T)$ to obtain an effective mass of $m^{*} \approx$ 0.022$m_{e}$ (see inset Fig. \ref{fig:2}b). In comparison the effective mass of the InSb QW ($m^{*} \approx$ 0.028$m_{e}$) was larger, confirming that alloying InSb with InAs indeed reduces the mass (see inset Fig.4b in the SI \cite{supplemental}). It should be noted that both masses exceed what is reported in literature for a similar InSb heterostructure, where $m^{*} \approx$ 0.017$m_{e}$ while the well width was comparable \cite{PhysRevResearch.2.033213}. We attribute this to the fact that asymmetric doping can skew the potential well, effectively reducing its width and resulting in a higher mass due to stronger confinement. In the upcoming section we will give a more comprehensive analysis of the confinement enhancement based on band structure considerations\par

\begin{figure}[t]
\centering
\includegraphics[width=8.4cm,height=7cm]{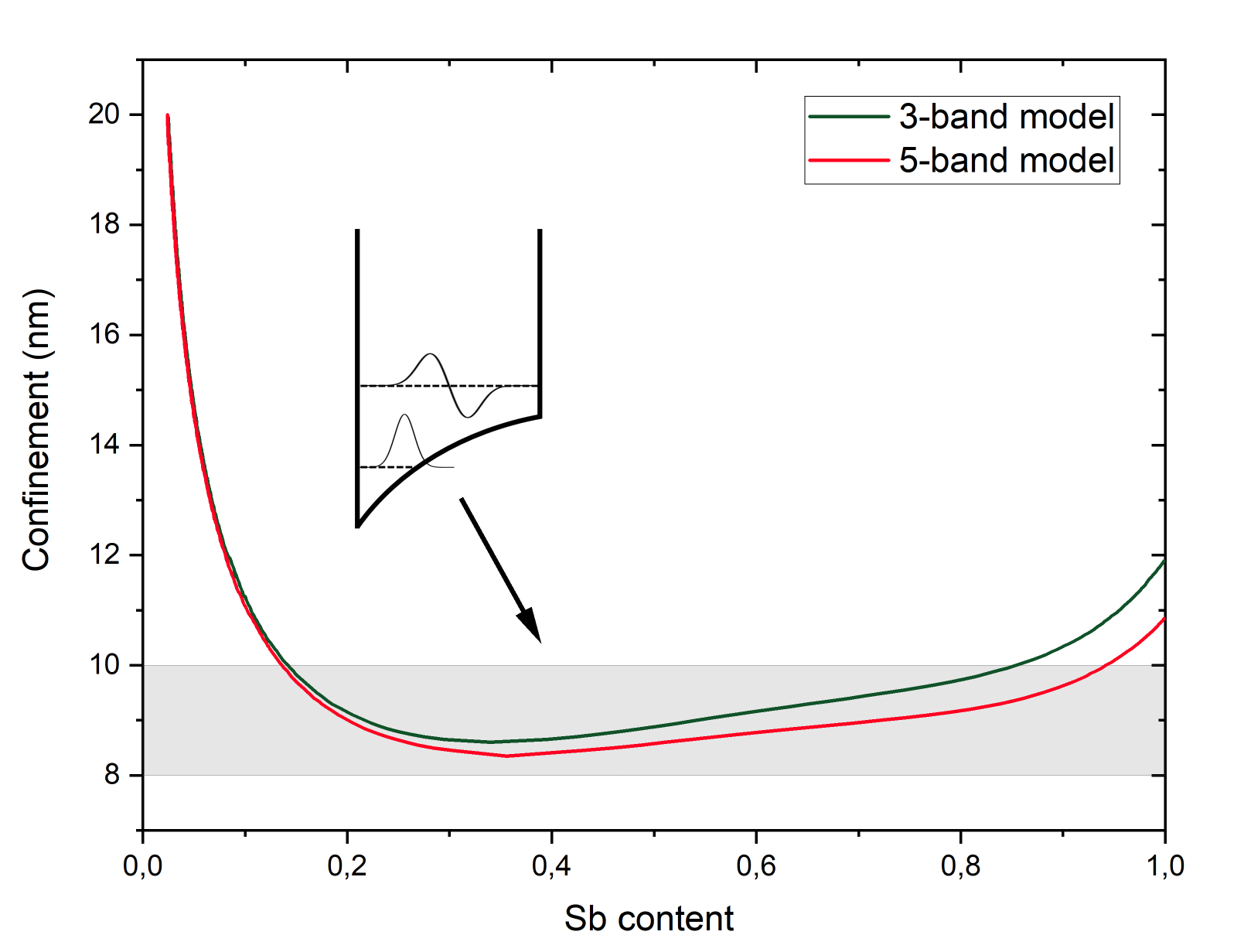}
\caption{Assuming an effective mass of 0.022$m_{e}$, we calculated the quantum well confinement with the three and five band \textit{\textbf{k $\cdot$ p}} theory as function of the Sb content in InAs$_{1-x}$Sb$_{x}$. A schematic depiction of the QW potential in the shaded area was included, showing the stronger confinement.}
\label{fig:8}
\end{figure}

Alternatively, one can also extract the effective mass from FIR transmission experiments, by tracking the CR as function of magnetic field. As shown in Fig. \ref{fig:7}a we observe clear dips in the transmittance, the signature of a CR, between 1-4 T before the sample becomes opaque due to the InSb phonon absorption band. The position of the resonance was determined by fitting the spectra with a Lorentzian and was plotted against magnetic field in Fig. \ref{fig:7}b. By performing a linear fit of $E_{CR}$ vs $B$, we can determine the effective mass, which was $m^{*} \approx$ 0.022$m_{e}$. This result is in perfect agreement with the mass obtained from the temperature dependent SdH measurements. \par 
It should also be noted that the CRs are quite broad, considering that the width goes from 2.4 to 4.3 meV due to the interaction with the phonon band. These line widths correspond to a scattering time of the order of 0.1 ps, which is comparable to the Drude scattering time determined below. For comparison, the width of the resonances in the InSb QW only increases from 1 to 1.7 meV within this field range (see Fig. 6a in the SI \cite{supplemental}), corroborating our assumption that alloying introduces additional short range disorder as the anion atoms will be randomly distributed over the available sublattice sites, which is in line with the lower mobility. \par 

\begin{figure}[t]
\centering
\includegraphics[width=8.4cm,height=7cm]{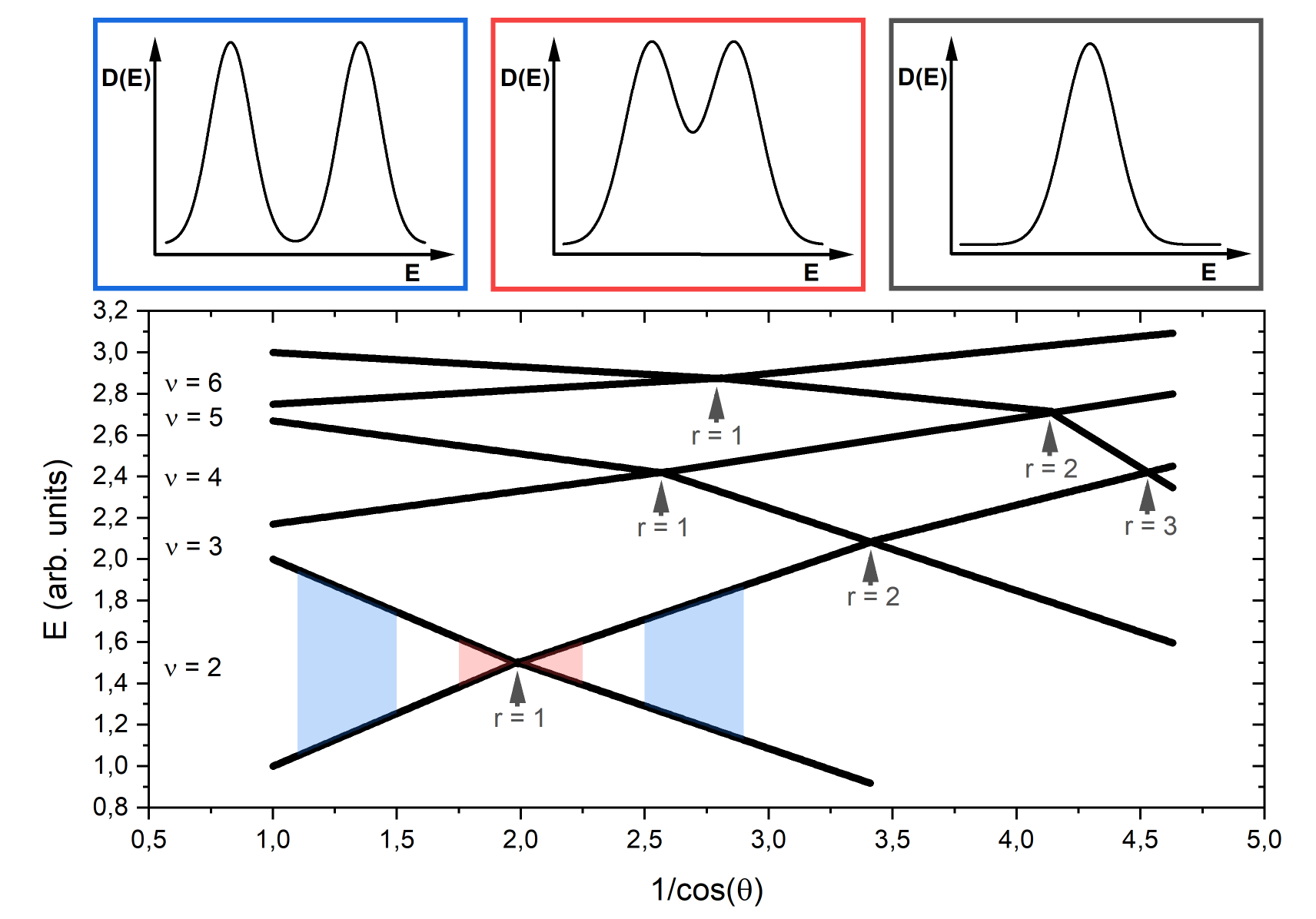}
\caption{Schematic representation of the principle behind a coincidence measurement, where one can see the opening and closing of the energy gap between even and odd Landau states. By including the effect of $g$-factor enhancement, one can see that coincidences (grey arrows) no longer occur at the same fields. Above the Landau fan diagram one can see cross-sections of the density-of-states at different angles, which correspond to the shaded areas between the Landau levels.}
\label{fig:4}
\end{figure}

In Fig. \ref{fig:9} we extract the quantum lifetime $\tau_{q}$ from the slope of the linear fit of 1/$B$ vs ln($\frac{\Delta\rho_{xx}}{\rho_{0}}A_{T}(B,T))$. Using the effective mass obtained above, $\tau_{q}$ was found to be 0.027 ps and was constant up to at least 20 K (see Fig. \ref{fig:9}). The Drude scattering time $\tau_{D}$ = 0.53 ps, which is determined with the classical Drude model, has the same order of magnitude as the scattering time extracted from the line width of the CRs. Knowing $\tau_{D}$ enables us to calculate the Dingle ratio as $\tau_{D}/\tau_{q} \approx$ 20. The fact that this value is significantly larger than 1, means that long-range potential fluctuations are the dominant scattering mechanism \cite{ihn2009semiconductor}. \par

\begin{figure*}[t]
\centering
\includegraphics[width=16cm,height=9cm]{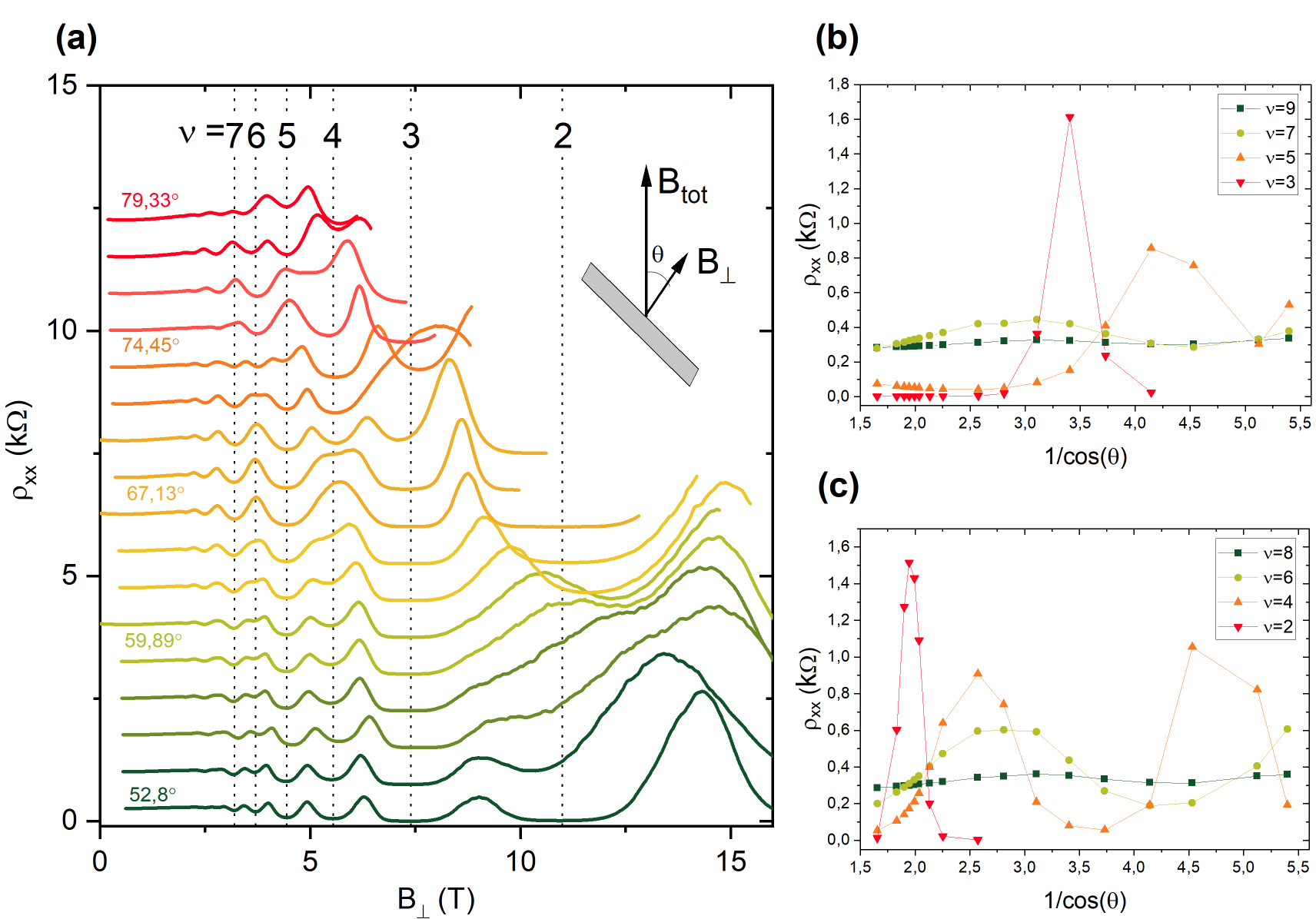}
\caption{(a) Angle dependence of $\rho_{xx}$ in an InAsSb QW, where the angle was determined by aligning the $\nu$ = 5 minima. The curves have a constant offset of 750 $\Omega$. (b) and (c) Show $\rho_{xx}$ as a function of $\theta$ for odd and even filling factors, respectively.}
\label{fig:3}
\end{figure*}

As the temperature dependence of the longitudinal conductivity $\sigma_{xx}$ at integer filling factors follows $\sigma_{xx} \propto exp(-E_{act}/2k_{B}T)$, we can determine the activation energy $E_{act,\nu}$ of the individual filling factors. From the linear fits presented in the inset of Figure \ref{fig:1}a, we derive that $E_{act,(1,2,4,6)}$ are 28.2, 23.4, 7.4 and 3.2 meV, respectively. By realizing that for odd filling factors $E_{act,odd} = g^{*}\mu_{B}B - \Gamma$ and neglecting the effect of the LL broadening $\Gamma$, we can calculate the effective $g$-factor. Therefore, at filling factor $\nu$ = 1 we have $g^{*} \sim$ 21.9. Plotting $E_{act,even}$ as a function of $B$ and fitting this linear relation with the expression $E_{act,even} = \hbar\omega_{c} - g^{*}\mu_{B}B - \Gamma$, gives us an alternative method to determine the effective mass as well as a rough estimate of the broadening (see Fig. 5a in the SI \cite{supplemental}). Due to the fact that we underestimate $g^{*}$, the effective mass $m^{*} \approx$ 0.029$m_{e}$ is slightly larger than the one obtained from the temperature dependence of the SdH oscillations and FIR transmission. The LL broadening $\Gamma$, given by the offset of $E_{act,even}$ vs $B$, is found to be 7.4 meV. Here, we did not take into account any magnetic field dependence of $\Gamma$. \par
Next, we compare the activation energies derived above, with the temperature where the QHE disappears. In literature there are reports that the thermal energy at which the Hall resistivity is still quantized is roughly eight times larger than the energy gap between the levels \cite{PhysRevB.93.125308}. Based on this phenomenological condition, we would expect the QHE to survive up to 41 or 34 K for $\nu$ = 1 and 2, respectively. Our magneto-transport measurements show that quantized quantum Hall plateaus survive up to 40 K, whereas signatures of the effect itself are present till 60 K (see Fig. \ref{fig:1}a). Consequently, we can conclude that the robustness of the QHE is nicely in line with the activation energies of the lower filling factors.

\subsection{Determination of confinement}

Next, with calculations based on multiband \textit{\textbf{k $\cdot$ p}} theory, we quantify confinement due to potential well skewing. First of all, we have to realize that the effective band gap $E_{g}$ will be increased due to the confinement of the QW, which in the infinite potential approximation amounts to $E_{con} = \pi^{2}\hbar^{2}/2L^{2}m^{*}_{bulk}$, where $L$ is the QW width and $m^{*}_{bulk}$ the effective mass of the bulk material. By only including the first order \textit{\textbf{k $\cdot$ p}} terms, one uses a three-level model neglecting all other bands except the conduction and valence band, giving us an expression as in Ref.~\onlinecite{ihn2009semiconductor},
\begin{equation} 
    \frac{m_{e}}{m^{*}} = 1 + \frac{2m_{e}P^{2}}{\hbar^{2}(E_{g} + E_{con})}.
\end{equation}
If all \textit{\textbf{k $\cdot$ p}} terms of the five closest bands at the $\Gamma$ point including the spin orbit interaction are taking into account, one can express the effective mass as follows,
\begin{equation}
\begin{split}
    \frac{m_{e}}{m^{*}} & = 1 + \frac{1}{3}\frac{2m_{e}P^{2}}{\hbar^{2}}\biggl(\frac{2}{E_{g} + E_{con}} + \frac{1}{E_{g} + \Delta_{0} + E_{con}}\biggl) \\
    &- \frac{1}{3}\frac{2m_{e}P'^{2}}{\hbar^{2}}\biggl(\frac{2}{E'_{g} - E_{g} + \Delta'_{0}} + \frac{1}{E'_{g} - E_{g}}\biggl),
\end{split}
\end{equation}

where $E_{g}$, $E'_{g}$, $\Delta_{0}$, $\Delta'_{0}$, $P$ and $P'$ are band-edge parameters which can be found in literature \cite{ihn2009semiconductor,vurgaftman2001band}. All band-edge parameters have a quadratic dependence on the alloy composition, which deviates from the linear interpolation between two binary compounds by defining a so-called Bowing parameter \cite{vurgaftman2001band}. \par
Figure \ref{fig:8} shows the dependence of the confinement on Sb content $x$ when $m^{*}$ = 0.022$m_{e}$ for both the 3- and 5-band model. Here we assumed that the Bowing parameter for $E_{g}$ is similar to that of $E'_{g}$, $P$ and $P'$. In both cases, we see an abrupt increase in confinement as $x$ is increased from 0 to 0.1, after which the dispersion starts to flatten off and eventually decreases again. EDX scans, shown in Fig. 2 of the SI \cite{supplemental}, QW confirm the presence of both As and Sb, which in conjunction with the fairly flat dispersion around the desired Sb content of 0.62, suggest that the confinement length will most likely lie between 8 and 10 nm. This rather strong confinement would also explain the relatively high effective masses we have reported earlier on. \par

\begin{figure}[t]
\centering
\includegraphics[width=8.4cm,height=7cm]{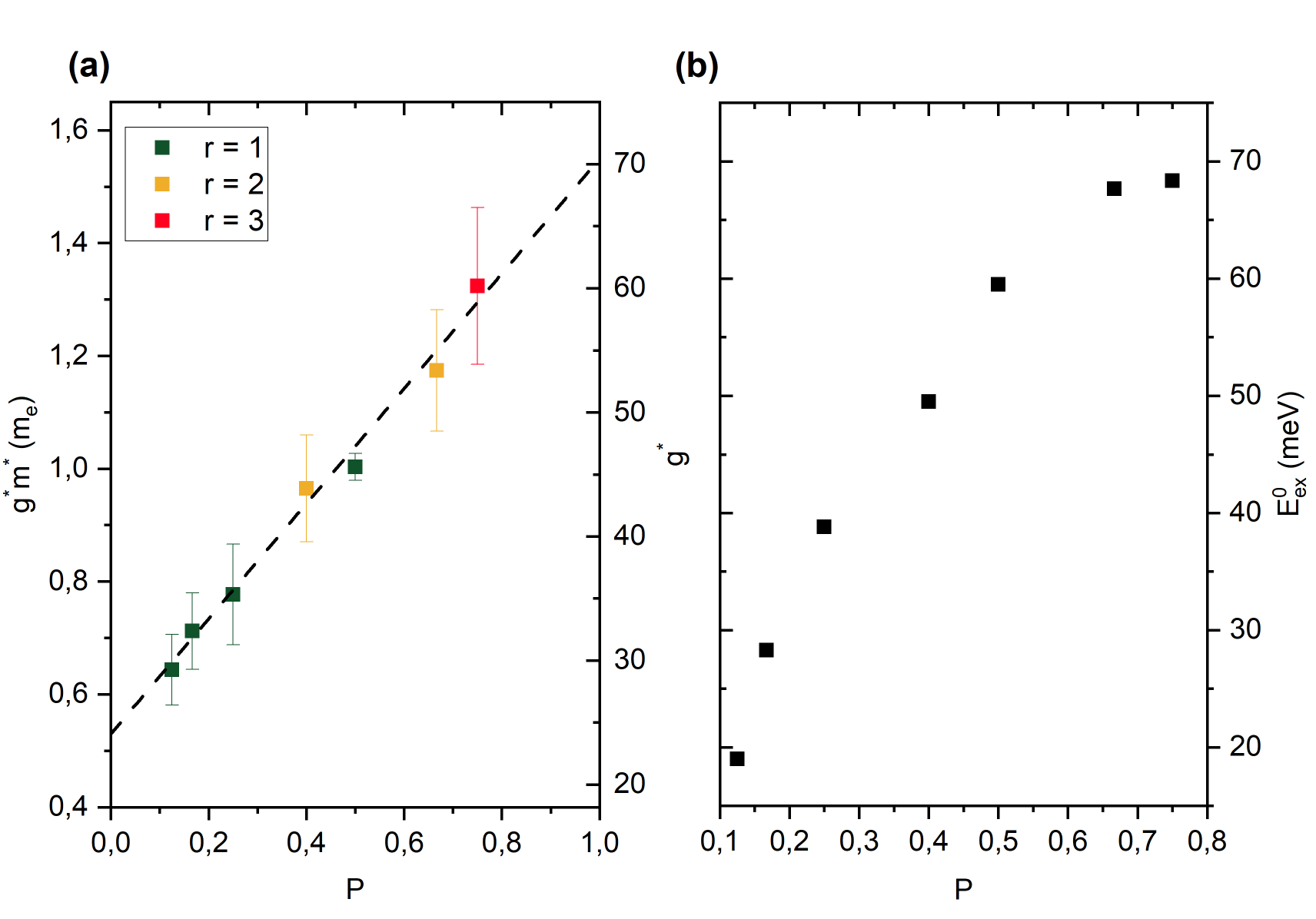}
\caption{(a) $g^{*}m^{*}$ as a function of $P$. Using the effective mass obtained by from the mass analysis, $m^{*} = 0.022m_{e}$, one can determine the effective $g$-factor. The corresponding $g^{*}$ values are shown on the right-hand side axis. (b) $E_{ex}^{0}$ as function of $P$. The exchange parameter was extracted from the linear fit of $g^{*}$.}
\label{fig:6}
\end{figure}

\subsection{Coincidence measurement}
In this section we will discuss the interaction-induced enhancement of the $g$-factor as function of spin polarization. Due to this enhancement, the $g$-factor can become much larger than predicted by \textit{\textbf{k $\cdot$ p}} theory in the single particle picture \cite{PhysRevResearch.4.013039}. To investigate this effect, we have performed coincidence measurements. The principle of a coincidence measurement, is to change the ratio between the LL spacing and the Zeeman energy by varying the angle $\theta$ between the direction of sample normal and total applied magnetic field $B_{tot}$ (see inset Fig. \ref{fig:3}a). Here we utilize the fact that the LL energy $\hbar\omega_{c}=\frac{\hbar eB_{\perp}}{m^{*}}$ is proportional to the perpendicular field component $B_{\perp}$, whereas the Zeeman energy depends on $B_{tot}$. Let us define the parameter $r$, which is the ratio between these two energies, i.e., $r = g^{*}\mu_{B}B_{tot}/\hbar\omega_{c}$. Since $B_{\perp} = cos(\theta)B_{tot}$ and using the definition of the Bohr magneton ($\mu_{B} = e\hbar/2m_{e}$), we can rewrite this expression as $rcos(\theta) = g^{*}m^{*}/2m_{e}$. For the case $r$ = 1, the Zeeman splitting $g^{*}\mu_{B}B_{tot}$ equals the LL spacing $\hbar\omega_{c}$, meaning that for even filling factors the Landau states with opposite spins have the same energy, closing the energy gap between them.  A similar reasoning holds for $r$ = 2, but now the spin splitting is twice as large as the LL spacing, resulting in the suppression of the odd filling factors. Thus, the longitudinal resistivity minima appear only at odd or even filling factors for $r$ = 1 and $r$ = 2, respectively. The opening and closing of the energy gaps is schematically depicted in Fig. \ref{fig:4}. \par 

\begin{figure}[t]
\centering
\includegraphics[width=8.4cm,height=7cm]{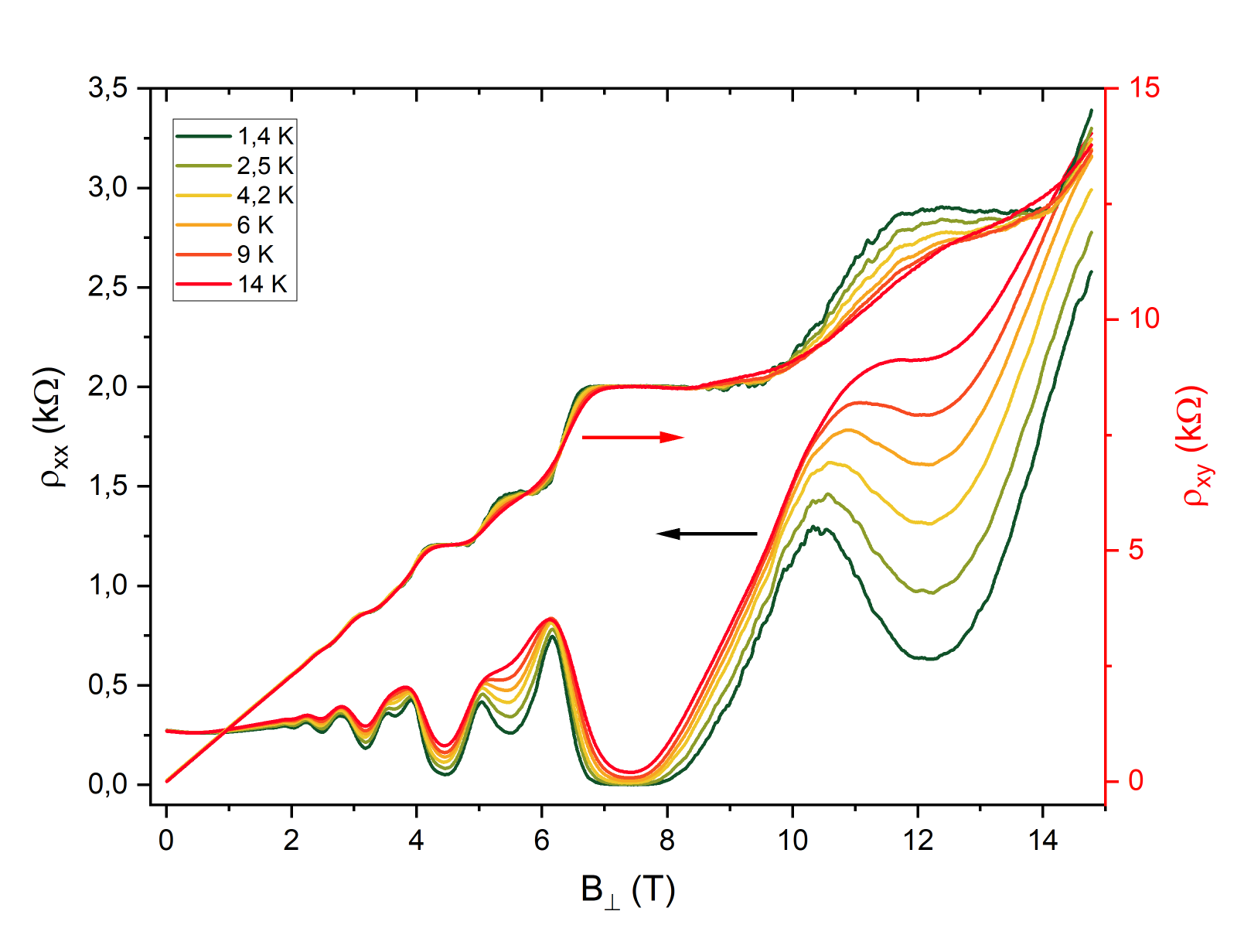}
\caption{Magnetic field dependence of $\rho_{xx}$ and $\rho_{xy}$ in an InAsSb QW at different temperatures when $\theta$ = 60.48$^{\circ}$. This angle correspond to $r$ = 1 for $\nu$ = 2.} 
\label{fig:5}
\end{figure}

Figure \ref{fig:3}a shows the result of our coincidence measurement, where we measured $\rho_{xx}$ as a function of $B_{\perp}$ at angles ranging from $\theta$ = 53$^{\circ}$ to 79$^{\circ}$. At each filling factor, indicated by the dashed lines, we monitor the evolution of $\rho_{xx}$ as the angle is increased. The cross-sections of these traces are plotted in Fig. \ref{fig:3}b and \ref{fig:3}c for odd and even filling factors, respectively. It becomes apparent from the position of the maxima that the $g$-factor is not universal for the individual filling factors, as the coincidences occur at different angles. With the expression for $r$ we can relate these coincidences to $g^{*}m^{*}$, which is proportional to the spin susceptibility $\chi = g^{*}m^{*}/2\pi\hbar$ in 2D systems \cite{PhysRevLett.90.056805}, or equivalently to $g^{*}$ after normalizing it with $m^{*}$. Figure \ref{fig:6} shows the dependence of $g^{*}m^{*}$ on the spin polarization $P$, defined as $P = r/\nu$. The data displays a monotonic increase with spin polarization and $g^{*}$ can become as large as 70 when the system is fully spin polarized at $P$ = 1. Such enhancement of the $g$-factor was also reported in Ref. \cite{PhysRevResearch.4.013039}, where an undoped InSb QW was investigated. The linear increase of $g^{*}$ can be expressed as $g^{*} = g_{0}^{*} + E_{ex}^{0}P/\mu_{B}B_{tot}$, where $g_{0}^{*}$ is the bare $g$-factor and $E_{ex}^{0}$ is the exchange parameter \cite{PhysRevB.84.121407}. $E_{ex}^{0}$ initially increases as one increases $P$, but starts to saturate around 70 meV at the highest spin polarization. For comparison, in graphene the exchange parameter amounts to 11.2 meV at 10 T, whereas in the InAsSb QW it is 28.3 meV at 10.5 T, which points towards stronger electron-electron interactions in this system \cite{PhysRevB.84.121407}. At the same time we see that the value for the $g$-factor obtained from the activation energies when the field was perpendicular to the sample, should be treated as the bare $g$-factor considering that $g^{*}$ = 24.1 at $P$ = 0. Next, we will compare this result to \textit{\textbf{k $\cdot$ p}} theory, where $g^{*}$ is given by 

\begin{equation}
\begin{split}
    g_{0}^{*} & = 2 - \frac{2}{3}\frac{2m_{e}P^{2}}{\hbar^{2}}\biggl(\frac{1}{E_{g} + E_{con}} + \frac{1}{E_{g} + \Delta_{0} + E_{con}}\biggl) \\
    &+ \frac{2}{3}\frac{2m_{e}P'^{2}}{\hbar^{2}}\biggl(\frac{1}{E'_{g} - E_{g}} + \frac{1}{E'_{g} - E_{g} + \Delta'_{0}}\biggl),
\end{split}
\end{equation}

such as in Ref.~\onlinecite{ihn2009semiconductor}. From this expression we get $g^{*}$ = 39.9 if we take $L$ = 8 nm and $x$ = 0.62. Like in the previous section, the band-edge parameters were determined using a quadratic interpolation between the two binary compounds. The obtained $g$-factor is larger than the experimentally determined value. This could be explained by the uncertainty in the two independent parameters, as $g^{*}$ strongly depends on the values used for the calculation. \par 

Finally, we return to our discussion of the QHE observed at rather large temperatures and propose scenarios to increase this temperature even further. The large splitting between spin levels within one Landau level, will significantly reduce the activation energy of even filling factors and with it also the temperature where the QHE disappears. The robustness of the QHE can be increased by placing the sample under the coincidence condition $r$ = 1 for filling factor $\nu$ = 2, where by definition the energy gap for $\nu$ = 1 is equal to the full LL spacing. In this particular case the energy gap is 58.4 meV, which would suggest that the QHE disappears at $\sim$85 K if we take into account the condition $E_{act} \gg k_{B}T$. In our case the carrier concentration is too high to reach this condition, but in theory this method can be used to increase the robustness. To illustrate this principle, we can again determine the activation energies of the QHE at $r$ = 1 for filling factor $\nu$ = 2. But one can already see in Fig. \ref{fig:5} that the energy gaps between the Landau states for even filling factors are significantly reduced, whereas the odd filling have become more robust. $E_{act,(2,3,4,5,6,7)}$ are found to be 0.54, 10.4, 0.42, 3.9, 0.082 and 1.1 meV, respectively. From the linear fit of $E_{act,odd}$ vs $B$ in Fig. 5b of the SI we get $g^{*} \approx$ 37.8 and $\Gamma \approx$ 5.85 meV \cite{supplemental}. The same can be done for the even filling factors, giving us $m^{*} \approx$ 0.024$m_{e}$, which is in good agreement with the value extracted from the temperature dependent SdH measurement (see Fig. 5c in the SI \cite{supplemental}). 

\section{Conclusions}
In summary, we have shown that by alloying InSb with InAs we can reduce the effective mass. The masses extracted from the CRs are in good agreement with those obtained from the temperature dependent SdH measurements. The fact that the masses are larger than the bulk values can be ascribed to a strong 2D confinement. The QHE is observed up to 60 K, which is in line with the obtained activation energies. The robustness of the QHE could be drastically improved be increasing the confinement length and bringing the sample into coincidence at $r$ = 1 for filling factor $\nu$ = 2. Though, for the latter we would need to decrease the carrier concentration to still be able to reach $\nu$ = 1 with our current setup. Also, an interaction-induced enhancement of the $g$-factor was observed with the coincidence method. At the largest spin polarization, $g$ can be as large as 60. From the linear fit of the $g$-factor as function of spin polarization we can extract values for the exchange parameter, which are significantly larger than those reported for graphene at comparable magnetic fields. To conclude, we have presented important proof-of-principle experiments on InAsSb QWs, a promising system with the potential of realising a low mass, high g-factor system.

\begin{acknowledgments}
This work was supported by HFML-RU/NWO-I, member of the European Magnetic Field Laboratory (EMFL). It is part of the research programme “HFML-FELIX: a unique research infrastructure in the Netherlands. Matter under extreme conditions of intense infrared radiation and high magnetic fields” with project number 184.034.022 financed by the Dutch Research Council (NWO). We also acknowledge financial support from the Swiss National Science Foundation (SNSF) and the National Center of Competence in Science ”QSIT-Quantum Science and Technology”.
\end{acknowledgments}

\begin{figure*}[t]
    \centering
    \includegraphics[width=14cm,height=7cm]{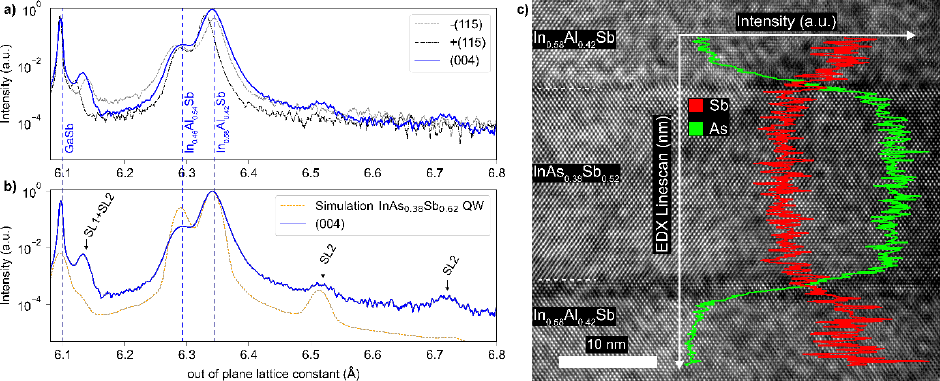}
    \caption{(a) The x-ray diffraction measurements depict symmetric (004), asymmetric (115-), and (115+) scans. These X-ray rocking curves reveal the diffraction peaks of the GaSb substrate, including the AlSb/GaSb superlattice (SL1), the In$_{46}$Al$_{54}$Sb/In$_{58}$Al$_{42}$Sb buffer, and satellite peaks of the In$_{46}$Al$_{54}$Sb/In$_{58}$Al$_{42}$Sb superlattice (SL2). The peak corresponding to the InAsSb quantum well (QW) either coincides as expected with the In$_{58}$Al$_{42}$Sb Layer peak or is relatively weak. The asymmetric rocking curves reveal that the In$_{58}$Al$_{42}$Sb layer is slightly strained (with 95$\%$ relaxation). (b) A simulation of the symmetric (004) X-ray rocking curve of the InAs$_{0.38}$Sb$_{0.62}$ QW heterostructure is presented. (c) A transmission electron microscopy (TEM) image demonstrates the InAsSb QW surrounded by InAlSb wells. Qualitative evidence of As atoms in the QW layer is substantiated by the energy-dispersive X-ray (EDX) linescan.}
    \label{fig:10}
\end{figure*}

\section{Supplemental information}

\subsection{Sample characterization}
\begin{figure}[b]
    \centering
    \includegraphics[width=8.4cm,height=7cm]{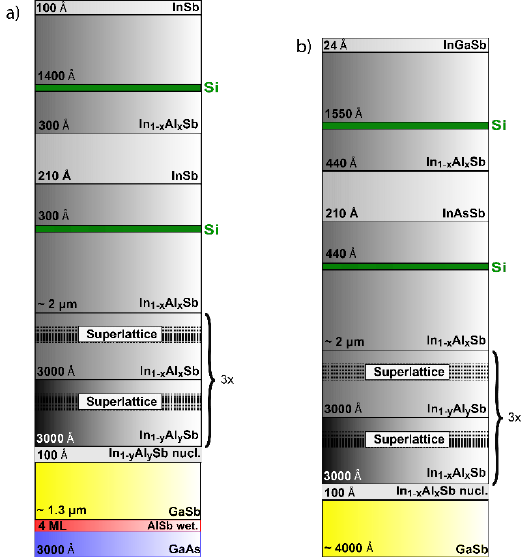}
    \caption{Layer stack of the InSb QW Structure (a) (with x=0.1,y=0.3) and the InAsSb QW structure (b) (with x=0.42, y=0.54).}
    \label{fig:11}
\end{figure}

The layer stack of InSb, along with the InAsSb quantum wells (QWs), is illustrated in Fig \ref{fig:11}. Further insights into the characteristics of the InSb QW can be found in the works of Christian Lehner \cite{20.500.11850/336239,PhysRevMaterials.2.054601}. Utilizing these investigations, adjustments were made to the buffer structure of the InAsSb QW sample in order to achieve a nearly lattice-matched InAlSb barrier enveloping the QW. For the intended InAs$_{0.38}$Sb$_{0.62}$ QW, this adaptation resulted in an In$_{0.46}$Al$_{0.54}$Sb/In$_{0.58}$Al$_{0.42}$Sb buffer structure.\par

The fabrication of the samples took place using a molecular beam epitaxy (MBE) Veeco Gen II system, with As and Sb being supplied via valved cracker sources. The temperatures of the As and Sb crackers were set at 750 $^{\circ}$C and 700 $^{\circ}$C, respectively. A pyrometer was employed to measure the substrate temperature. The growth rates of group \RomanNumeralCaps{3} elements were determined through the analysis of Reflection High-Energy Electron Diffraction (RHEED) oscillations. The \RomanNumeralCaps{5}/\RomanNumeralCaps{3} ratios were then deduced from these growth rates of group \RomanNumeralCaps{5} and group \RomanNumeralCaps{3} elements.\par

The growth rate of As (Sb) was derived by tracking As (or Sb) oscillations using the following method: Initially, the As (Sb) flux was terminated, and approximately 10 monolayers (MLs) of Ga were deposited onto the surface of GaAs (GaSb) at a rate of 0.5 $\angstrom$/s and a temperature of 570 $^{\circ}$C (480 $^{\circ}$C). Immediately thereafter, As (Sb) was introduced at a selected valve setting, resulting in distinct and well-defined RHEED oscillations. For the InAsSb QW, an As (Sb) flux corresponding to a GaAs (GaSb) growth rate of 0.7 $\angstrom$/s (4.8 $\angstrom$/s) was employed. The In$_{0.46}$Al$_{0.54}$Sb/In$_{0.58}$Al$_{0.42}$Sb buffer and the InAsSb QW were grown with a constant InAs (InSb) growth rate of 1.4 $\angstrom$/s (1.7 $\angstrom$/s) at a substrate temperature of 430 $^{\circ}$C.\par
The composition of InAs$_{x}$Sb$_{1-x}$ layers is influenced not only by the appropriate As/Sb flux ratio but also by other growth conditions, including the growth temperature and the specific As and Sb species from the valved cracker sources \cite{20.500.11850/336239}. To ensure accurate composition control, thicker InAs$_{x}$Sb$_{1-x}$ layers on an In$_{1-y}$Al$_{y}$Sb/In$_{1-x}$Al$_{x}$Sb buffer were utilized as a reference, and their composition was calibrated using (004) X-ray diffraction measurements.\par

\begin{figure}[t]
    \centering
    \includegraphics[width=8.4cm,height=7cm]{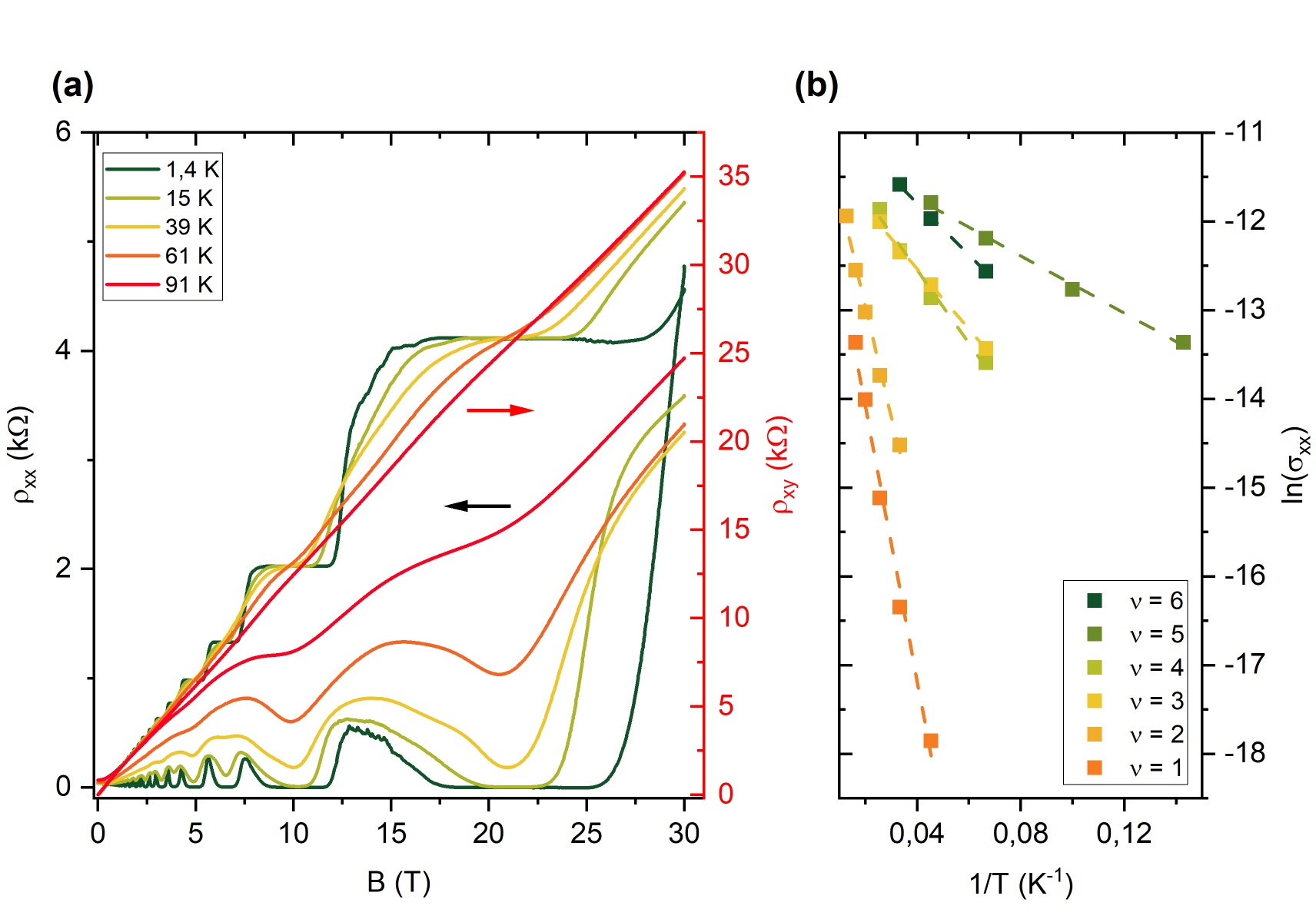}
    \caption{Magnetic field dependence of $\rho_{xx}$ and $\rho_{xy}$ in an InSb QW at different temperatures. (b) Temperature dependence of $\sigma_{xx}$ at multiple filling factors. Using the relation $\sigma_{xx} \propto exp(-E_{act}/2k_{B}T)$, we extracted the activation energies from the linear fits.}
    \label{fig:12}
\end{figure}

\begin{figure}[t]
    \centering
    \includegraphics[width=8.4cm,height=7cm]{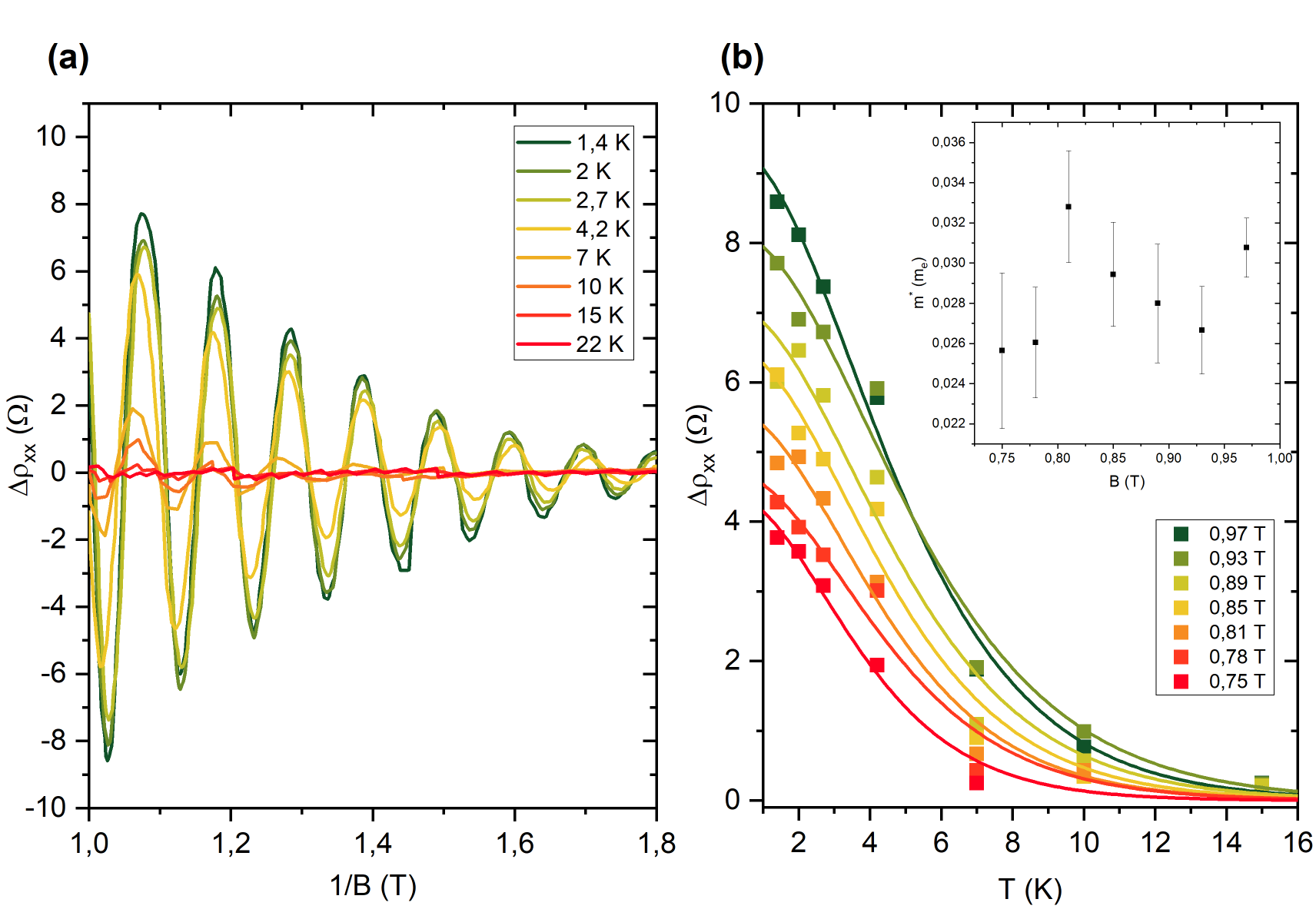}
    \caption{(a) Temperature dependence of the SdH oscillations in an InSb QW after subtraction of a magnetoresistance background. Shown is only the field range were no spin splitting is observed. (b) Temperature dependence of quantum oscillation amplitudes at different magnetic fields. The data was fitted with $A_{T}(B,T)$, which describes the thermal damping of quantum oscillations, to obtain the effective mass (see inset).}
    \label{fig:13}
\end{figure}

Figure \ref{fig:10}a displays X-ray rocking curves of the final InAsSb QW heterostructure in the symmetric (004) direction along with the complementary directions (115-) and (115+). The anticipated peak of the InAs$_{0.38}$Sb$_{0.62}$ QW either coincides with the peak of the In$_{0.58}$Al$_{0.42}$Sb layer or exhibits a weak signal due to the relatively thin nature of the QW in comparison to the buffer structure. A simulation of the X-ray rocking curve for this heterostructure (Fig. \ref{fig:10}b) confirms the validity of this assumption. The faint satellite maxima emerge due to In$_{0.46}$Al$_{0.54}$Sb/In$_{0.58}$Al$_{0.42}$Sb superlattices in the buffer (SL2) and an AlSb/GaSb superlattice in the GaSb layer (SL1). The InAs$_{0.38}$Sb$_{0.62}$ QW is examined via TEM and EDX. The EDX linescan qualitatively confirms the presence of As atoms in the QW layer.\par
A qualitative assessment of the strain in the QW and the buffer can be inferred from measurements using the asymmetric diffraction geometry. The non-coincident maxima positions of In$_{0.58}$Al$_{0.42}$Sb indicate a slightly strained pseudomorphically grown QW (~95$\%$ relaxation). Conversely, the maxima of In$_{0.46}$Al$_{0.54}$Sb overlap, indicating relaxation in these layers.

\subsection{Effective mass analysis InSb}
Figure \ref{fig:12}a shows the magnetic field dependence of a 405 $\times$ 25 $\mu m^{2}$ section of the InSb QW Hall bar at different temperatures. At the lowest temperatures we can resolve $\nu$ = 42 and $\nu$ = 22 in $\rho_{xx}$ and $\rho_{xy}$, respectively. This is in good agreement with the condition $\mu B \gg 1$, which predicts that the onset of SdH oscillations will occur at 0.3 T, whereas in the $\rho_{xx}$ trace at 1.4 K they start around 0.44 T. \par
After subtracting a magneto-resistance background $\rho_{0}$, we get the actual quantum oscillation amplitude $\Delta\rho_{xx}$ (see Fig. \ref{fig:13}a). At this point we again monitor the oscillation amplitude as function of temperature and fit the resulting points with $A_{T}(B,T)$ to determine the value of the effective mass, which was $m^{*} \approx$ 0.028$m_{e}$ (see inset Fig. \ref{fig:13}b). \par

The activation energy, which was determined from the longitudinal resistivity $\rho_{xx}$ at integer filling factors, for $\nu$ = 1 and $\nu$ = 2 are 26.8 and 21.4 meV, respectively (see Fig. \ref{fig:12}b). Based on these activation energies we would expect the QHE to survive up to 39 or 31 K for $\nu$ = 1 and $\nu$ = 2, respectively. Our magneto-transport measurements show that quantized quantum Hall plateaus survive up to 40 K (see Fig. \ref{fig:12}a). Consequently, we can conclude that the robustness of the QHE is nicely in line with the activation energies of the lower filling factors. A direct comparison between the $\rho_{xy}$ traces at 60 K for both types of QWs reveals that due to the smaller mass the QHE is more pronounced in InAsSb. 

\subsection{Field-dependence activation energy InAsSb}
\begin{figure}[t]
    \centering
    \includegraphics[width=8.4cm,height=7cm]{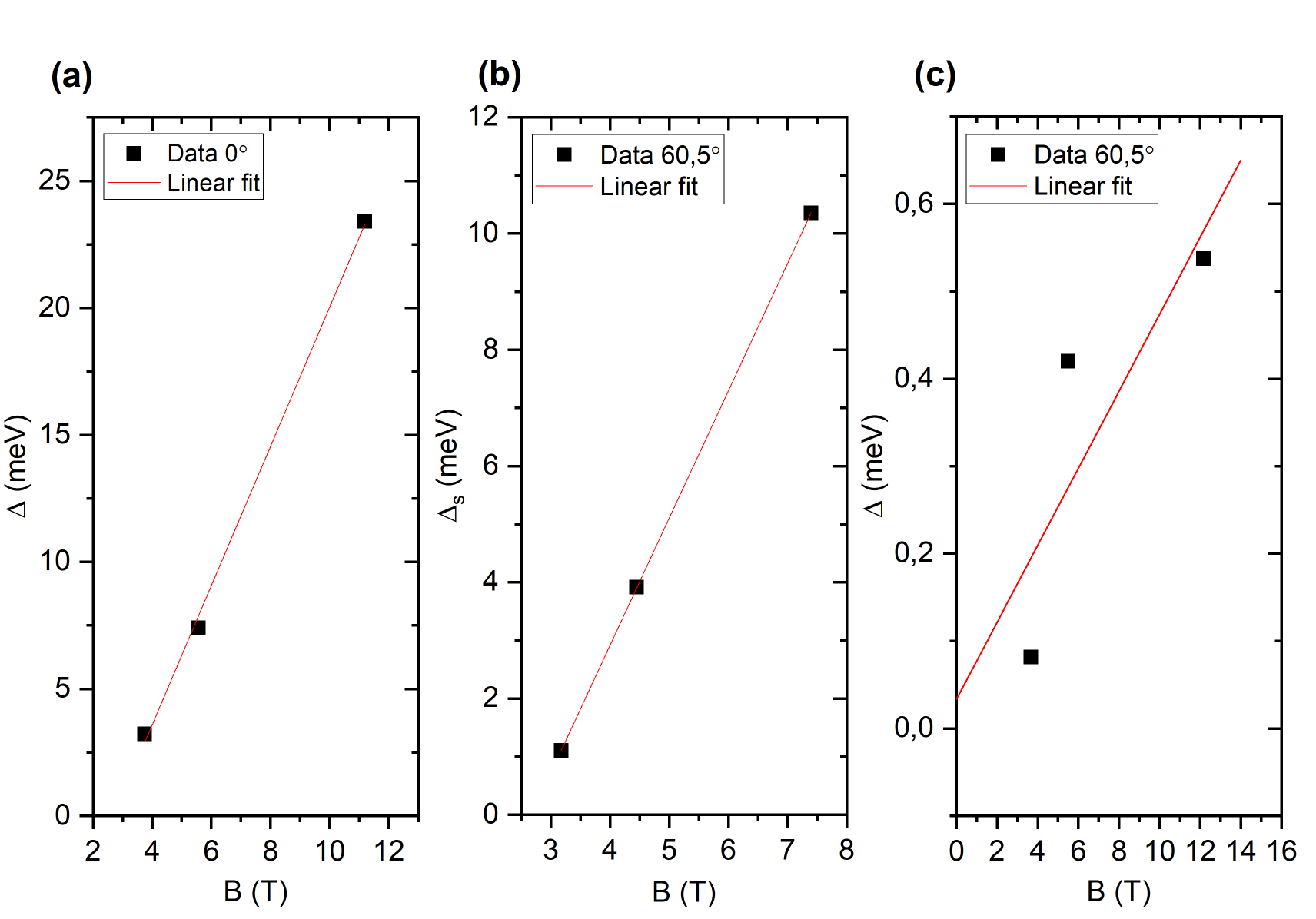}
    \caption{(a) Field-dependence of the activation energy in an InAsSb QW for even filling factors at $\theta$ = 0$^{\circ}$. (b) and (c) Field-dependence of the activation energy in an InAsSb QW for odd and even filling factors at $\theta$ = 60.5$^{\circ}$, respectively. The red lines represent the linear fits used to determine $m^{*}$ and $g^{*}$.}
    \label{fig:14}
\end{figure}

As the activation energy of both even and odd filling factors have a linear dependence on field, if we neglect the field dependence of LL broadening, we can extract values for $m^{*}$ and $g^{*}$ from a linear fit of the data. Figure \ref{fig:14}a displays a fit of the even filling factors activation energies ($\nu$ = 2, 4 and 6) when the sample is oriented perpendicular to the magnetic field. With the $g^{*}$ value calculated from $E_{act,1}$, we can determine the effective mass from the slope and we find that $m^{*} \approx$ 0.029$m_{e}$. The same can be done for $\theta$ = 60.5$^{\circ}$, corresponding to the coincidence condition $r$ = 1 for $\nu$ = 2. From the magnetic field dependence of the odd and even filling factor in Fig. \ref{fig:14}b and c, respectively, we get $g^{*} \approx$ 37.8 and $m^{*} \approx$ 0.024$m_{e}$. 

\begin{figure}[b]
    \centering
    \includegraphics[width=8.4cm,height=7cm]{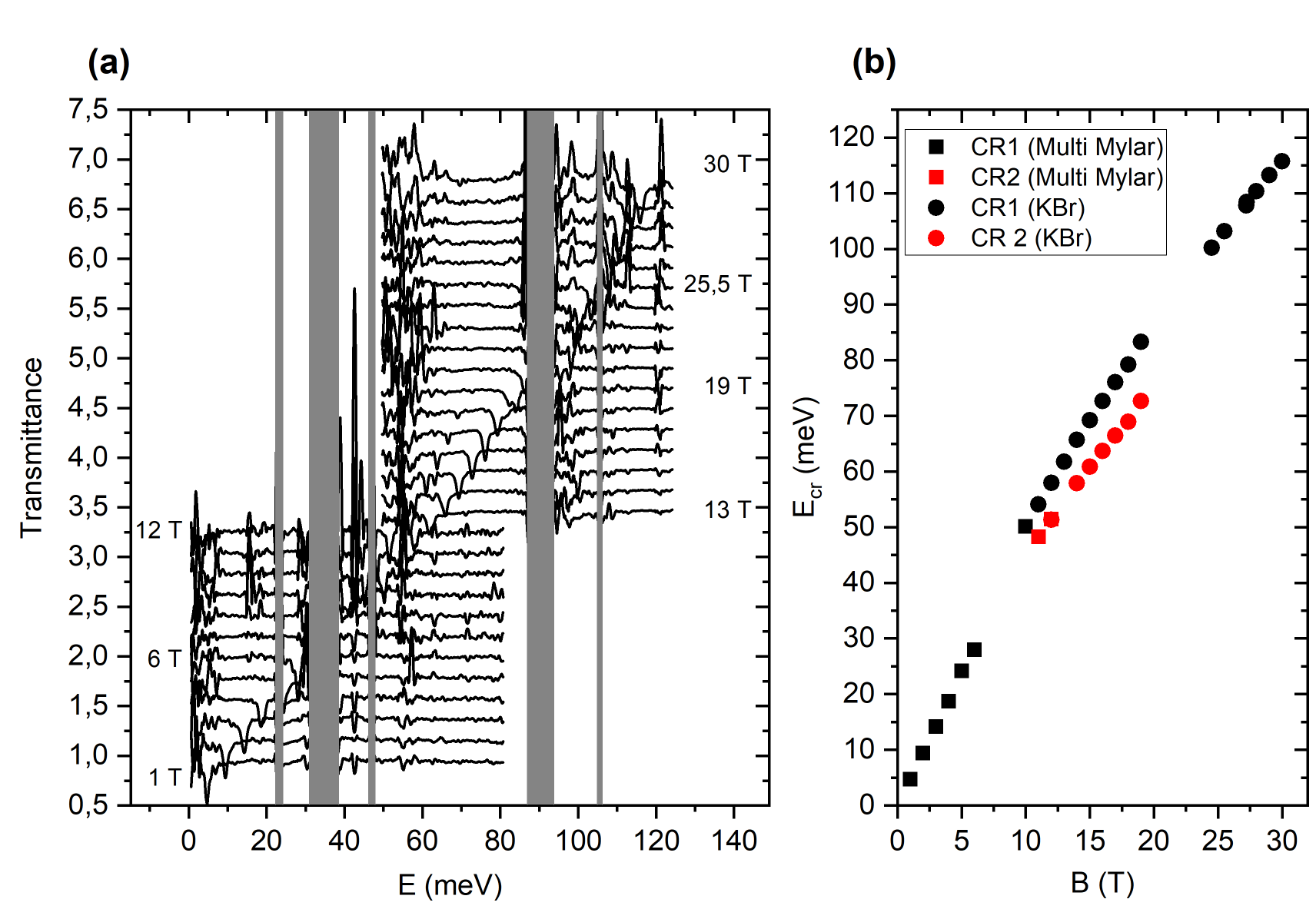}
    \caption{(a) Transmittance as a function of wavenumber in an InSb QW for different magnetic fields. The traces have a constant offset of 0.2. The phonon absorption bands have been grayed out. In the field range between 11-18 T there are two CR features corresponding to different LL transitions. (b) Cyclotron resonance energy as function of magnetic field. Here, we used a multilayered Mylar beamsplitter for $\tilde{\nu}$ = 5 - 650 cm$^{-1}$ and a KBr beamsplitter for $\tilde{\nu}$ = 400 - 1000 cm$^{-1}$.}
    \label{fig:15}
\end{figure}

\subsection{FIR transmission InSb}
The transmittance spectra contain CR features all the way up to 30 T (see Fig. \ref{fig:15}a). In the field range between 11-18 T there are two distinct CR dips present due to the non-equidistant LL spacing stemming from the non-parabolicity of the bands. This non-parabolicity can be clearly seen in the non-linear magnetic field dependence of the CR energy (see Fig. \ref{fig:15}b). As the transition probability for CRs depends on the occupancy of the Landau levels, which changes drastically with magnetic field as they are pushed through the Fermi level, certain transitions will only be observed within a specific field range. The effective mass can be determined from the low-field data in Fig. \ref{fig:15}b, where the dispersion is still linear, giving us $m^{*} \approx$ 0.025$m_{e}$. This value is noticeably smaller than the value extracted from the transport data. Such a discrepancy could originate from the fact that the FIR transmission measurements were performed with a different piece of the wafer that distinctive values for the effective mass. 
\pagebreak


%

\end{document}